\def\rfr#1{eq. (\ref{#1})}

\def\Rfr#1{Eq. (\ref{#1})}

\def\virg#1{``#1''}

\def\eqi{\begin{equation}}
\def\eqf{\end{equation}}
\def\eqia{\begin{eqnarray}}
\def\eqfa{\end{eqnarray}}
\def\rp#1#2{{#1\over#2}} \def\lb#1{\label{#1}}

\def\bds#1{\boldsymbol{#1}}

\documentclass{ws-ijmpd}

\begin{document}

\markboth{L. Iorio}
{\virg{Imprinting} in general relativity tests?}

%
\catchline{}{}{}{}{}
%

\title{\virg{IMPRINTING} IN GENERAL RELATIVITY TESTS?}

\author{LORENZO IORIO\footnote{Viale Unit\`{a} di Italia 68, 70125, Bari (BA), Italy}}

\address{INFN-Sezione di Pisa, Largo B. Pontecorvo 3\\
Pisa,  56127
Italy\\
lorenzo.iorio@libero.it}

\maketitle


\begin{abstract}
We investigate possible a-priori \virg{imprinting}  of general relativity itself on spaceraft-based tests of it. We deal with some performed or proposed time-delay ranging experiments in the Sun's gravitational field. The \virg{imprint} of general relativity on the Astronomical Unit and the solar gravitational constant $GM_{\odot}$, not solved for in the so far performed spacecraft-based time-delay tests, may induce an a-priori bias of the order of $10^{-6}$ in typical solar system ranging experiments aimed to measuring the space curvature PPN parameter $\gamma$. It is too small by one order of magnitude to be of concern for the performed Cassini experiment, but it would affect future planned or proposed tests aiming  to reach a $10^{-7}-10^{-9}$ accuracy in determining $\gamma$.
\end{abstract}


\section{General considerations}\lb{uno}
In many GTR tests several observations from space probes are confronted to predictions for them computed with given dynamical models. The relativistic effect to be tested is explicitly included in them, with one or more solve-for parameters $\{P\}$ accounting for it to be estimated in a least-square fashion, along with many other ones $\{K\}$ not directly pertaining GTR. Of crucial importance for interpreting such data analyses as genuine tests of GTR is to clarify how the numerical values of the models' parameters $\{F\}$  which have not been solved-for  have originally been obtained.
The point is that the standard data reduction procedure used for the original goals of the missions, now \virg{opportunistically} exploited for  GTR testing, may not be valid, in principle, for performing a truly unbiased, genuine check of GTR which is not a \virg{tautology}.
Indeed, if the primary task of a space-based mission is, for example, to reach a certain astronomical target with a given accuracy, the only thing that is important to this aim is that the dynamical models adopted to predict the probe's motion are accurate enough; this is usually quantitatively judged by inspecting the post-fit residuals of some directly measurable quantities like, e.g., ranges or range-rates. How the parameters $\{F\}$ entering the models have been obtained, i.e. their a-priori reference values, does not matter at all: the only important thing  is that the resulting fit of an existing set of  observations is good enough to minimize the observable's residuals.
Such  an approach may, in principle,  not be entirely adequate when the goal of the data analysis is testing a gravitational theory like GTR in an unambiguous, unbiased and self-consistent way. In this case, how the fixed parameters $\{F\}$  of the models have been obtained does, in fact, matter. Indeed, if one or more of them $\{I\}$ were previously obtained from different data of different bodies in such a way that they somehow retain a non-negligible a-priori \virg{imprint} of the same effect we are now interested in, their use may bias the current test just towards the desired outcome yielding, for example, a very high accuracy confirmation. In this cases, it would be more correct to use, if possible, values of such \virg{imprinted} parameters $\{I\}$ which have been obtained independently of the effect itself whose existence we are just testing in the present data analysis, even if the accuracy of such different values of the \virg{suspect} parameters $\{I\}$ was worse. Alternatively, if, for some reasons, such \virg{unbiased} values are not available, $\{I\}$ should be included, if possible, in the list of the solved-for parameters along with the one(s) $\{P\}$ accounting for the effect to be tested, and the resulting covariance matrix should be checked to inspect the correlations among them.
%
\section{Application to some specific cases}\lb{due}
To be more definite, let us look at the Cassini radio science test. In that case, the radiotechnical data of the spacecraft traveling to Saturn were contrasted with  a set of dynamical models by JPL of its motion and electromagnetic waves propagation in such a way that a correction $\Delta \gamma$ to the GTR-predicted value of the PPN parameter $\gamma$ was solved for, among other parameters, obtaining\cite{Ber03} \eqi\Delta\gamma\doteq|\gamma - 1|=(2.1\pm 2.3)\times 10^{-5};\eqf other authors got\cite{And04}
\eqi\Delta\gamma\doteq|\gamma - 1|=(-1.3\pm 5.2)\times 10^{-5}.\eqf
Now, a physical parameter which is crucial in such a test is the gravitational constant $GM_{\odot}$  of the Sun, which is the source of the relativistic time delay. It was not estimated \cite{Ber03,And04}, so that its numerical value was kept fixed to the standard reference figure of the JPL DE ephemerides. It does, in principle, contain an a-priori \virg{imprinting} by GTR itself through the same effect itself that was just tested with Cassini, in particular by $\gamma$ itself. Indeed, the numerical value of $GM_{\odot}$ comes from the fixed value of the defining Gaussian constant\footnote{See on the WEB http://ssd.jpl.nasa.gov/?constants.}
\eqi k = 0.01720209895\ {\rm au^{3/2}\ d^{-1}},\eqf and from the value of the Astronomical Unit\footnote{Here we will use au for the symbol of the Astronomical Unit, like m for the meter, while AU will denote its numerical value in m. Instead, d denotes both the symbol of \virg{day} and its numerical value in s.} in m, not estimated  in the Cassini tests,
\eqi {\rm AU}=1.49597870691\times 10^{11}\ (\pm 3)\ {\rm m}\eqf  through
\eqi GM_{\odot} = k^2\ {\rm AU}^3\ {\rm d}^{-2} = 1.32712440018\times 10^{20}\ (\pm 8\times 10^9)\ {\rm m^3}\ {\rm s^{-2}}.\lb{giemme}\eqf AU was, in fact, obtained just through a combination of radar ranging of Mercury, Venus, and Mars, laser ranging of the Moon (making use of light reflectors left on the lunar surface by Apollo astronauts), and timing of signals returned from spacecraft as they orbit or make close passes of objects in the solar system\cite{Sta04}; thus, as we will show below, it is affected in a non-negligible way, given the level of accuracy of the techniques adopted,  by GTR itself and, in particular, by $\gamma$  which enters the PPN expressions for the time delay and bending of traveling electromagnetic waves.
Thus, there exists, in principle, the possibility that the high-accuracy results of the Cassini radio science tests may retain an a-priori \virg{imprint} of GTR itself through $GM_{\odot}$ (and the Astronomical Unit as well).

Let us put our hypothesis on the test by making some specific calculations; for the sake of clarity, we will refer to the Cassini radio science tests, but the conclusions may be considered valid also for any of the many proposed $\gamma-$dedicated missions.
The GTR time delay experienced by electromagnetic waves propagating from point 1 to point 2 is
\eqi\Delta t = \rp{2R_g}{c}\ln\left(\rp{r_1 + r_2 + r_{12} + R_g}{r_1 + r_2 - r_{12} + R_g}\right),\lb{dela}\eqf
where $R_g\doteq 2GM_{\odot}/c^2$ is the Sun's Schwarzschild radius; $r_1$ is the heliocentric coordinate distance to point 1, $r_2$ is the heliocentric coordinate distance to point 2, and $r_{12}$ is the distance between the points 1 and 2. \Rfr{dela} is the expression actually used in the JPL'S Orbit Determination Program (ODP) used to analyze interplanetary ranging with planets and probes. In order to quantitatively evaluate the level of \virg{imprinting} by GTR itself in the used value of the Astronomical Unit, let us assume $r_1$ equal to the Earth-Sun distance and let us vary $r_2$ within 0.38 au and 1.5 au to account for the ranging to inner planets; the maximum effect occurs at the superior conjunction, i.e. when
${\bds{ n}}_1\approx -{\bds{n}}_2$, and $r_{12}\approx r_1 + r_2$. It turns out that $\Delta t_{\rm ranging}\approx 4\times 10^{-4}$ s, which is certainly not negligible with respect to the accuracy of the order of $10^{-8}$ s with which the light-time for\footnote{The value in km of the Astronomical Unit is obtained by measuring at a given epoch the distance between the Earth and a target body (a planet or a probe orbiting it)  by multiplying $c$ times the round trip travel time $\tau$ of electromagnetic waves sent from the Earth and reflected back by the target body, and confronting it with the distance, expressed in AU, between the Earth and the target body at the same epoch as predicted by some accurate dynamical ephemeris \cite{Sta04}.} 1 au $\tau_{\rm A}$ is actually measured (http://ssd.jpl.nasa.gov/?constants). As a consequence, the quantitative impact of the interplanetary ranging in the inner solar system to the determination of the Astronomical Unit is of the order
\eqi d{\rm AU}=c\Delta t_{\rm ranging} = 1.14291\times 10^5\ {\rm m},\lb{diau}\eqf not negligible with respect to the meter-level accuracy in measuring the Astronomical Unit; thus, $d$AU$/$AU$=8\times 10^{-7}$. Differentiating \rfr{giemme} with respect to au and \rfr{diau} yields
\eqi \rp{dGM_{\odot}}{GM_{\odot}}=2\times 10^{-6}.\lb{digiemme}\eqf Thus, we conclude that the technique adopted to determine the numerical values of the Astronomical Unit and of the Sun's $GM$ induced an a-priori \virg{imprint} of GTR on them of $8\times 10^{-7}$ and $2\times 10^{-6}$, respectively.

Let us apply this result to a typical radio science experiment in the solar system with $r_1$ fixed to the Earth-Sun distance. By writing $r_{1/2}= x_{1/2}$ au, with $x_{1/2}$ expressing distances in Astronomical Units, differentiation of \rfr{dela} with respect to au and $GM_{\odot}$, and \rfr{diau}-\rfr{digiemme} yield
an \virg{imprinting} effect of the order of\eqi \left.\rp{\delta(\Delta t)}{\Delta t}\right|_{\rm GTR} = 2\times 10^{-6}\eqf for $r_2$ up to tens\footnote{The Cassini test was performed with $r_2=7.43$ au \cite{Ber03}.} AU; it turns out that the largest contribution comes from $dGM_{\odot}$. It is too small by one order of magnitude with respect to the performed Cassini radio science tests, but it should be taken into account in the future, more accurate experiments whose expected accuracy is of the order of $10^{-7}-10^{-9}$, in the sense that the a-priori bias of GTR in the future determinations of deviations of $\gamma$ from unity will be as large as, or even larger than the effects one will to test, unless either $GM_{\odot}$ will be estimated as well along with $\gamma$ itself or a value obtained independently of it will be adopted.


\begin{thebibliography}{0}    

\bibitem{Ber03}
B. Bertotti, L. Iess and P. Tortora,  \textit{Nature} \textbf{425} (2003) 374.
%
\bibitem{And04}
J.D. Anderson, E.L. Lau and G. Giampieri,  Measurement of the PPN Parameter $\gamma$ with Radio Signals from the Cassini Spacecraft at X- and Ka-Bands,  in \textit{Proc. of the 22nd
Texas Symp. on Rel. Astrophys., Stanford, eConf C041213, 0305}, eds. P. Chen, E. Bloom, G. Madejski and V. Patrosian (SLAC, Stanford, 2004).
%
\bibitem{Sta04}
E.M. Standish,  The Astronomical Unit Now, in \textit{Transits of Venus: New Views of the Solar System and Galaxy
Proceedings IAU Colloquium No. 196, 2004}, ed. D.W. Kurz (Cambridge University Press, Cambridge, 2005), p. 163.
\end{thebibliography}
\end{document}